\documentstyle[11pt,twoside,jltp]{article}

\title{Massive Charged Strings in the Description of  
Vortex Ring Quantum Nucleation} 
\author{Uwe R. Fischer\thanks{e-mail: 
fischer@tat.physik.uni-tuebingen.de}
\address{Institut f\"ur Theoretische
Astrophysik, University of T\"ubingen, Germany}}
\runninghead{Uwe R. Fischer}{Massive Charged Strings in Vortex Ring Quantum 
Nucleation}
\begin{document}
\begin{abstract}
We demonstrate the way in which vortex rings in neutral superfluids 
are equivalent to massive, charged, elastic strings in an electromagnetic 
field defined locally on the string from a Kalb-Ramond gauge field. 
We argue that the action thus obtained describes an 
intermediate scale of vortex
motion with phonon fluctuations of the line between the incompressible 
hydrodynamic r\'egime and the microscopic
one dominated by roton emission and absorption. 
The formalism gives an accurate semiclassical picture 
of vortex string motion
and is of relevance 
for the description of vortex ring quantum nucleation 
in a perturbation theory of the purely incompressible case. 
\end{abstract}
\maketitle
\bigskip
We will discuss here in a condensed form
the motion of a slightly deformed vortex ring 
as a massive, elastic string object under the influence of the 
nondissipative 
Magnus force and clarify the requirements for 
the validity of this formulation.
 
In the hydrodynamic limit of a 2+1d superfluid there exists  
a direct correspondence between the genuinely
relativistic electron-positron pair interaction via photons in QED
and a `relativistic' point-vortex-point-antivortex pair interaction via 
phonons\cite{popov2,cyclo}.
To obtain the motion of a vortex line, 
we extend this analogy to the 3+1d case by introducing elastic energy
of the vortex {\it string}.   
The  `relativistic' result for vortex motion in 2+1d as well as 3+1d 
will break down long before the local vortex 
velocity reaches the velocity of sound because the hydrodynamic
treatment will cease to be valid.  The microscopic excitations 
and vortex structure will
come into play and the limiting velocity of vortex motion
will be of the order of the Landau critical velocity of roton creation $v_L$
rather than the sound velocity. 
To describe the failure of the structureless entity approximation of
points and strings one usually introduces the  
ultraviolet cut-off $\xi$
related to the core radius of the vortex and the microscopic
excitations of the superfluid, i.e. its many-body structure.
Using the notion {\it string} for a one-dimensional object
with an extension $\xi$ in the  normal direction to the actual string
location with only phonons as excitations of the superfluid 
amounts to the requirement that at any
point on the string the curvature is much less than $\xi^{-1}$ or 
equivalently that it moves locally much slower than the speed of sound
$c_s$.
We consequently use in the Lagrangian 
the non-`relativistic' limit for the motion of the vortex (whereas the
phonon background is necessarily `relativistic'). 
The local frequency of vortex motion will then be much slower
than a critical  $\omega_\xi \equiv c_s/\xi$.
Calculating $\omega_\xi$ with $\xi$ slightly above 
half the interparticle spacing in He II (the specific example of
superfluid we have in mind here), 
yields indeed a value of $\omega_\xi$ very close to the frequency of the
roton minimum. 

In what follows, $X^\mu (t,\sigma)$ designates the spacetime location of the
string (with $\sigma$ the arc length parameter),
$\dot X^\mu\equiv \partial X^\mu\! / \partial  t$, 
$X'^\mu\equiv \partial X^\mu\! /\partial \sigma$. The speed of sound
$c_s\equiv 1$ in the vortex self-action
is understood, except where $c_s$ is 
explicitly indicated for the convenience of the reader.
To generalize the 2+1d result to the 3+1d case we set up a 
local righthanded basis on the string given by the negative 
normal, tangent and binormal of the line:
$\vec e_1 (t,\sigma) = -\vec{X}''/|\vec X''|,\,{\vec X}',\,
\vec e_2 (t,\sigma) = \vec{X}'\times \vec e_1 \,.$
We first consider the dynamics of the vortex described by the vector $\vec Q$ 
lying in the plane spanned by $\vec e_1,\vec e_2$ 
coming from its local self-interaction.
Only in the case of small 
perturbations from an 
equilibrium configuration the equations of motion of 
vortices do obey a Hamiltonian structure \cite{fettquth}. In
particular, self-crossings of a line bending back on itself
have to be excluded.
For a three-dimensional superfluid such an equilibrium configuration 
is given by a circular vortex, for which 
${\vec X}'={\vec e}_\Phi,
{\vec e}_1=\vec{e}_R,\vec{e}_2=\vec{e}_Z$
and $Q^1=R,\, Q^2=Z$. 
Under this prescription of small perturbations of equilibrium, 
the vortex self-action 
is now written as a sum of its static part and a
wave-dynamical one, quadratic in derivatives of $\vec Q$:
\begin{equation}\label{self}
S_{\rm self}[\vec Q(t,\sigma)]=
-\int\!\int dt d\sigma\sqrt{\gamma(t,\sigma)}\left\{ M_0 
-\frac12 M_0\dot{\vec Q}{}^2 +\frac12\frac\alpha{\gamma}\, 
{\vec Q}'{}^2\right\}
\,.
\end{equation}
Here, $\sqrt{\gamma(t,\sigma)}\,d\sigma$ is the measure of the string's proper
length (for the circular one $\sqrt\gamma =R(t)$). 
The canonical way to obtain 
the static hydrodynamic kinetic mass $M_0$ of the vortex per unit
proper length 
is to calculate an effective action as a functional
of the vortex coordinate alone. 
\cite{popov2,duan,cyclo,wexler}.
The same result, though, is obtained simply by considering the vortex string
as a fundamental object in the superfluid.  
The mass $M_0$  
is then the hydrodynamic vortex energy per unit proper length 
in the local vortex rest frame (the renormalized string tension), 
divided by the sound speed squared. 
Reinstating $c_s$ for this Einsteinian mass-energy relation:  
\begin{equation}\label{Einstein}
 M_0 c_s^2= 
\frac{m\rho_0\kappa^2}{4\pi}
\left[\ln \left(\frac{8R_c}{\xi e^C}\right)
\right]\,.
\end{equation}
The infrared cut-off in the logarithm is in the static limit 
equal to the mean distance of line elements, i.e. proportional to the 
local curvature radius $R_c=|\vec {X}''(t,\sigma)|^{-1}\gg\xi$. 
The constant $C$ depends on the
core structure 
and has order unity.
For `relativistic' vortex motions approaching 
the frequency $\omega_\xi$ 
the kinetic mass will depend on the position on the string
respectively the wavevector of its oscillations
\cite{cyclo,wexler}.
In general, the cut-off related elastic coefficient 
$\alpha$ (parameterized by the longitudinal  
as well as the transversal string core structure)
is also a function of $t,\sigma$.
We do not consider short wavelength perturbations, 
where this dependence becomes significant, and take $\alpha\equiv M_0$.
This specific choice in the elastic energy of the string  
is related to a  
cut-off in the localized self-induction approximation of classical
hydrodynamics \cite{schwarz1}. It corresponds to the assumption
that in the long wavelength limit we are using, only massless
sound excitations propagating along the string can survive 
and accounts for the fact that $\alpha\equiv 
M_0 c_s^2$ must remain finite in the incompressible limit 
$c_s\rightarrow\infty$. 

The linear coupling of the vorticity to the 
background gauge potential which we will add to (\ref{self})
represents the Magnus force term in the action \cite{gradwohl,zee}. 
It is a topological coupling, involving no local property of the
string, which in its most general, relativistic 
form reads:
\begin{equation}\label{Magnus}
S_M
=m\rho_0 \kappa\int\! d^2\zeta\, b_{\mu\nu} \dot X^\mu X'^\nu
\,.
\end{equation}
We represent the linear coupling of the background flow to the vorticity
by using the dual transformation to the antisymmetric gauge
tensor $b_{\mu\nu}$, first expounded in its relation to string
interactions and superfluids in
now-classic papers \cite{kalb,lundregge}. It relates the only dynamical 
degree of freedom of the order parameter we
consider, the phase $\theta$, to $b_{\mu\nu}$ and its field strength
$H_{\alpha\mu\nu}$:
\begin{equation}\label{dual}
\frac{\hbar}m\partial^\gamma\theta\,\epsilon_{\gamma\alpha\mu\nu}\,
=v^\gamma\epsilon_{\gamma\alpha\mu\nu}
\equiv 
H_{\alpha\mu\nu} =  b_{\mu\nu,\alpha} + b_{\alpha\mu,\nu} +
b_{\nu\alpha,\mu}\,.
\end{equation}  
The self-action (\ref{self}) with $\lambda = M_0$ could have been
obtained from (\ref{Magnus}) if we included in 
the Goldstone part of $b_{\mu\nu}$ the contribution of the 
vortex itself and integrated out the transverse phonon fluctuations 
of the line up to the relevant infrared and ultraviolet cut-offs. 
Here, $b_{\mu\nu}$ is solely 
the background flow gauge potential with wavevector below the  
cut-off $k_0\approx e^C/8R_c$ in (\ref{Einstein}) \cite{popov2}:
\begin{equation}\label{background}
b_{\mu\nu}(x)\equiv \frac1{(2\pi)^4}\int_{k<k_0} \!\!d^4 k\, 
\tilde b_{\mu\nu}(k) \exp [ik_\mu x^\mu]\,. 
\end{equation}

The Magnus force on a string is equivalent to a local form of the Lorentz force
on a particle.  
This becomes apparent if we define  
the local `electromagnetic' 3-potential and 3-current density via
\begin{equation}\label{adef}
(m\rho_0)^{-1/2}a_\mu\equiv b_{\mu\sigma}\equiv b_{\mu\nu} X'^\nu\;,\quad
(m\rho_0)^{-1/2}j^\mu\equiv \omega^{\mu\sigma} \equiv\omega^{\mu\nu} X'_\nu\, ,
\end{equation}
i.e. project the components of $b_{\mu\nu}$ respectively $\omega^{\mu\nu}$
on the local 
vortex axis, according to (\ref{Magnus}). 
In an arrangement of cylindrical symmetry, fulfilled by a ring vortex
($\sigma=\Phi$),
the 3-potential is the vector $a^\mu = (a^0,a^r,a^z)
=(m\rho_0)^{1/2}(-b_{0\phi},b_{r\phi},b_{z\phi})$.
The component $b_{0\phi}$ 
equals for stationary flows Stokes' stream function \cite{milne-t}. 
We identified $q\equiv (m\rho_0)^{1/2}\kappa $ with the `charge' 
of the vortex.
The `electric' and `magnetic' fields in the Magnus (Lorentz) force law 
resulting from the linear coupling of the vortex matter
coordinate to the local gauge field are represented by an analogous
projection
\begin{eqnarray}
F_{\mu\nu}\equiv (m\rho_0)^{1/2}H_{\mu\nu\beta}X'^\beta=
(m\rho_0)^{1/2} \,
\epsilon_{\mu\nu\alpha\beta}v^\alpha X'^\beta\,, 
\nonumber\\
\vec E =  
(m\rho_0)^{1/2} \vec v  \times \vec X' \; ,\quad  
\vec B =  
(m\rho_0)^{1/2}  v_0 \vec X'\,,
\end{eqnarray}
where we used the background flow 
$v_\mu = (\hbar /m) \partial_\mu \theta$ 
from the dual transformation (\ref{dual}). In a Galilei invariant
condensed matter system like He II, $v^0=1$ applies (the velocity of
light $c\equiv 1$).

Summing (\ref{self}) and (\ref{Magnus}), we obtain the total vortex action:
\begin{equation}\label{action}
S_V=
\int\!\!\int dt d\sigma\left\{-\sqrt\gamma M_0 
\left(1-
\frac12 \dot{\vec Q}{}^2 +\frac1{2\gamma} {\vec Q}'{}^2\right)
-q a^0
+ q {\vec a}\cdot\dot{\vec Q}
\right\}\,,
\end{equation}
and stress again that it is useful as long as we
are able to neglect retardation effects in the motion of the vortex
string, i.e. as long as the vortex moves non-`relativistically'. 
The fact that only linear and quadratic powers of $\dot{Q}_A$ 
appear in the Lagrangian for non-`relativistic' motions
can equivalently be understood as the
condition of adiabaticity for the vortex motion \cite{wexler}: 
If the effective action is written as a power series in $\dot{Q}_A$,  
the terms in the action (\ref{action}) are the first two.   
The canonical vortex momenta 
for the action (\ref{action}) are given by the expression 
\begin{equation}\label{Pcanon}
\vec P (t,\sigma) =
\vec{e}_A \frac{\delta L}{\delta \dot Q_A(t,\sigma)}=\sqrt\gamma M_0 
\dot {\vec Q}(t,\sigma) + q \vec a 
\, .
\end{equation} 
The Hamiltonian of the vortex 
is equivalent to that of a continous string sequence of 
particles enumerated by
$\sigma$, bound together by the elastic term $(1/2) M_0\vec Q'^2$,
which is subject to an `electromagnetic' 3-potential $a^\mu$
locally defined at the position of the string from the external gauge field
$b_{\mu\nu}$ in (\ref{adef}):  
$$
H_V[\vec Q,\vec P]
=  \int\! d\sigma \left\{\sqrt\gamma\, \left[M_0  
+ \frac 1{2\gamma M_0} \left( \vec P
-q\vec a
\right)^2+\frac{M_0}{2\gamma}\,
\vec Q'^2\right] +\frac12 q a_C^0 +q a_{\rm v_s}^0
\right\}
$$
$$
\hspace*{-14em}=
\int d\sigma \left[\sqrt\gamma M_0 +\frac12
qa_C^0+q a^0_{\rm v_s}\right]
$$
\begin{equation}\label{HV}
\hspace*{3em}+
\frac12 m\rho_0\lambda \int d\sigma \sqrt\gamma\,\frac{\bar\gamma}{\gamma}
\left[ \left(({\vec P -q\vec a})/{M_0}\right)^2+ \vec
Q'^2\right]
\equiv H_0+H_\lambda\,.
\end{equation}
We used the Coulomb (transverse) gauge condition 
div$\,\vec a=0\,\Leftrightarrow \,
b_{\sigma A}{}^{;A}=0$\,
and defined $a^0\equiv a_C^0+a^0_{\rm v_s}$, i.e. separated off  
the Coulomb and background velocity 
parts of the scalar potential. This gives
the correct factor of 1/2 in the Coulomb interaction energy of 
the vortex with other 
vortices contained in the
background part of the energy.
In the second part of expression (\ref{HV}) 
we introduced a scale $\bar R_c $
characterizing typical curvature radii of
vortices under consideration.
We defined $\bar\gamma\equiv \gamma (\bar R_c)$
and neglected
the weak dependence of $\lambda\equiv M_0/m\rho_0\bar\gamma$\, on
$\sigma$. 
For a ring vortex with radius $R$ the quantity 
$ \lambda =(\kappa^2/4\pi R^2)\ln (R/\xi e^C)\approx
(\xi /R)^2\ln(R/\xi e^C)/\pi $, where use was made of the approximate 
equality $\kappa \approx c_s 2\xi $. 
Thus the parameter $\lambda$ gives an
approximate measure (that is, up to logarithmic renormalization) of the 
inverse number of particles $1/N$ 
contained in a disc of radius
$\bar R_c$. 

If we wish to describe in the framework of our formalism
the nucleation of the string as a quantum
object, the canonical commutation
relations are to be imposed:
$$
[Q_A(\sigma),P_B(\sigma')]= i\hbar \delta_{AB}\delta(\sigma-\sigma')\;,
$$
\begin{equation}
[Q_A(\sigma), Q_B(\sigma')]  =  [P_A(\sigma),P_B(\sigma')]=0\,.
\label{canoncomm}
\end{equation}
It should be strongly emphasized  
that we are dealing here with a vortex
string which we have given an effective mass.
Were the compressibility zero and hence $M_0=0$, we
could not impose that canonical coordinates and momenta in 
the directions perpendicular to the string commute because 
they are then mutually canonically conjugated up to a 
factor\cite{volovik,fettquth}. 
If we solve the static background equation of constant `magnetic' field
$\vec B= $ rot\,$\vec a = -(m\rho_0)^{1/2}{\vec e}_\Phi$ 
in the Coulomb gauge, we have the possible gauge choice 
$a^z=1/2(m\rho_0)^{1/2}R^2$.
This gauge choice is such as to make the integrated expression 
(\ref{Pcanon}) coincide with the Kelvin momentum of the vortex
and prevents the imposition of (\ref{canoncomm}). Any equivalent 
choice, like the isotropic one $a^z= 1/3 (m\rho_0)^{1/2} R^2,\,
a^r= - 1/3 (m\rho_0)^{1/2}R Z$, will do the same.   

The Hamiltonian (\ref{HV}) has been split 
into a part $H_0$ belonging to a zero mass vortex without elastic energy
and a perturbation part $H_{\lambda}$ involving the wave term.
The 
Hamiltonian $H_0$ 
has been used \cite{volovik} to compute the quantum nucleation probabilities 
for vortex half ring formation in the presence of a half sphere at a boundary.
The part $H_{\lambda} $ 
represents a small correction 
and a perturbation theory based on 
the separation of $H_0$ and $H_{\lambda}$ 
can be constructed to incorporate the dynamics of the massive, elastic
vortex string in the calculation of such nucleation probabilities.

The microscopic examination of vortex ring 
quantum nucleation in the context of a U(1) field theory 
in analogy to pair creation in QED
\cite{davis3,kaolee}, in which compressibility and 
vortex mass play a dominant role,
suffers from the inadequacy of
the field theory in a real superfluid on scales of the coherence length 
respectively for velocities approaching the speed of sound. 
In contrast, the formalism developed here gives an
accurate idea of
what happens dynamically if the microscopic domain in a real
superfluid is approached from above.  
The equations of motion 
following from the action (\ref{action}) 
will be relevant for the
description of vortex ring quantum nucleation in an intermediate
region between incompressible hydrodynamics and the microscopic
domain because they are strictly valid in this semiclassical region.
It is thus possible to calculate the corrections the 
effective mass of a vortex string 
induces in the intrinsic nucleation probabilities 
of quantized vortices at the absolute zero of temperature
in He II.   
Our model should be verifiable with  
the outcome of experiments on intrinsic vortex nucleation in this superfluid
\cite{ihasAV},
where the plateau of the critical velocity temperature dependence below about 
150 mK is usually associated with the onset of a temperature
independent quantum r\'egime of nucleation.   
\label{cyl}
\section*{ACKNOWLEDGMENTS}
I would like to thank Grisha Volovik for stimulating discussions,
especially the initial one at the monastery of Saint Hugues. Fruitful 
conversations with Nils Schopohl are gratefully acknowledged.
This research work has been supported by a grant of 
the LGFG Baden-W\"urttemberg. 

\end{document}